\documentclass[pra,showpacs,preprintnumbers,amsmath,amssymb,twocolumn]{revtex4}

\usepackage{bm}
\usepackage{amsmath}
\usepackage[dvips]{graphicx}
\newcommand{\del}[1]{}

\begin{document}
\bibliographystyle{apsrev}

\title{The $g$ factor of an electron or muon bound by an arbitrary central potential}

\author{S.G. Karshenboim$^{(a,b)}$\footnote{Electronic address: sek@mpq.mpg.de}, R.N. Lee$^{(c)}$\footnote{Electronic address: R.N.Lee@inp.nsk.su}, A. I. Milstein$^{(c)}$\footnote{Electronic address: A.I.Milstein@inp.nsk.su}}

\address{$^{(a)}$D. I. Mendeleev Institute for
Metrology (VNIIM), St. Petersburg 198005, Russia\\
$^{(b)}$Max-Planck-Institut f\"ur Quantenoptik, 85748 Garching,
Germany\\
$^{(c)}$ Budker Institute of Nuclear Physics,
630090 Novosibirsk, Russia.}

\date{\today}

\begin{abstract}
We consider the $g$ factor of a spin-$1/2$ particle (electron or
muon) bound by an arbitrary central field. We present an approach
which allows one to express the relativistic $g$ factor in terms of
the binding energy. We derive the general expression for the
correction to the $g$ factor caused by a deviation of the central
potential from the Coulomb one. As the application of this method,
we consider the corrections to the $g$ factor due to the finite
nuclear size, including vacuum polarization radiative correction.
The effect of the anomalous magnetic moment is also taken into
account.
\end{abstract}

\maketitle

\section{Introduction}

Study of energy levels of atomic electron (muon) has a long history.
Various relativistic and radiative corrections have been obtained in
analytic or semi-analytic form taking into account a deviation of
the potential from the Coulomb one. Recently interest to an accurate
theory of the $g$ factor increased. Accurate theoretical approach
needs to take into account relativistic and radiative corrections.
Here we consider a case when such corrections correspond to some
modification of a central potential. Such a problem is actual
because of the Uehling potential, which is responsible for a
dominant QED correction in muonic atoms, finite-nuclear-size effects
and some others. The results are presented in a fully relativistic
approach, i.e. exact in $Z\alpha$ for the Coulomb interaction and
its modifications.

The methods of calculation of energy levels have been successfully
developed over the decades. The energy levels can be calculated with
high accuracy by solving certain equations analytically or
numerically, or using the perturbation theory with respect to the
corresponding correction to the hamiltonian. Calculation of the $g$
factor is a more complicated problem. In the traditional approach to
the calculation of the correction to the $g$ factor due to
modification of the potential $\delta V$, one considers both the
magnetic field and $\delta V$ as a perturbation, thus starting with
the second order of the perturbation theory. The numerical
approaches are also not easy to apply because they usually are much
more accurate in determination of the energy than the wave function.
However, it is the latter which is needed to calculate the $g$
factor.

In this paper, we develop a framework which allows one to express
the $g$ factor of a Dirac particle bound in arbitrary central
potential via the binding energy. For a small deviation of this
potential from the Coulomb one, we derive the general expression for
the first correction to the $g$ factor due to this deviation. The
results are valid both for muonic and electronic atoms.

\section{General relations}

In the weak homogenous magnetic field ${\bm B}$, the correction to
energy levels of a Dirac particle  bound by a central potential
reads
\begin{eqnarray}
 \Delta E &=& e \int d^3r \,\overline{\Psi}({\bm r})\bigl({\bm
\gamma}\cdot{\bf A} \bigr) \Psi({\bm r})\nonumber\\
&=&\frac{e\bm B\cdot \langle\bm J\rangle}{2m}\,g \label{eq:TM0}\,,
\end{eqnarray}
where
\begin{equation}\label{defAB}
{\bf A}=\frac{1}{2}\bigl[{\bm B}\times{\bf r} \bigr]\;.
\end{equation}
The relativistic units in which $\hbar=c=1$ are applied throughout
the paper; the charge of the bound particle is $-e$, $e$ is a charge
of a proton, $e^2=\alpha=1/137$ is the fine structure constant.

Writing the Dirac wave function $\psi({\bm r})$ in the form
\begin{equation} \label{Dirac}
\psi({\bm r})=
\begin{pmatrix}
f_1(r)\Omega\\
if_2(r)\widetilde{\Omega}
\end{pmatrix}
\,,
\end{equation}
where $\Omega$ is the spherical spinor \cite{BLP} with the angular
momentum $J$ and orbital momentum $L$, $\widetilde{\Omega}=-({\bm
\sigma}\cdot{\bm n})\Omega$, we obtain for the $g$ factor of a
bound Dirac particle
\begin{equation}\label{gf1f2}
g=\frac{2m\kappa}{j(j+1)}\,\int{dr}\,r^3 f_1(r)\, f_2(r)\;,
\end{equation}
$\kappa=j+1/2$ for $l>j$ and $\kappa=-(j+1/2)$ for $l<j$. The
integral in the right-hand side of Eq.~(\ref{gf1f2}) can be
presented as follows (see, e.g., \cite{rose})
\begin{equation}\label{identity}
\int{dr}\,r^3 \,f_1\, f_2=-\frac{1}{4m}\left[1-2\kappa\int{dr}\,r^2
\bigl(f_1^2-f_2^2\bigr) \right]\;.
\end{equation}

This formula is valid for the Dirac equation with any central
potential. Eq. (\ref{gf1f2}) has been known for a while (see, e.g.,
\cite{rose}), but, to the best of our knowledge, it was never
applied to a non-Coulomb problems.

Using the identity
\begin{equation}\label{g0}
\int{dr}\,r^2 \bigl(f_1^2-f_2^2\bigr)=\langle \gamma_0\rangle\;,
\end{equation}
we have
\begin{equation}\label{gfactor}
g= -\frac{\kappa[1-2\kappa \langle \gamma_0\rangle]}{2j(j+1)}\;.
\end{equation}

Assuming that the potential $V(r)$ is independent of the mass of
the bound particle, we find
\begin{equation}\label{g0HE}
\langle \gamma_0\rangle = \bigg\langle \frac{\partial H} {\partial
m}\bigg\rangle= \frac{\partial E} {\partial m}\;,
\end{equation}
where $H$ is the Dirac hamiltonian $H=\gamma_0\,\bigl({\bm
\gamma}\cdot{\bm p}\bigr) + \gamma_0 m + V(r)$.

Thus we arrive at an equation, which expresses the $g$ factor of the
state via its binding energy
\begin{equation}\label{important1}
g=-\frac{\kappa}{2j(j+1)}\, \left[1-2\kappa\frac{\partial E}
{\partial m}\right]\;.
\end{equation}
This equation can be used even in the case when a potential
substantially differs from the Coulomb one, as for heavy muonic
atoms. In many cases, Eq. (\ref{important1}) essentially simplifies
calculation of the corrections to the $g$ factor of a Dirac
particle. In particular, it can be applied to such problems as
finite-nuclear-size effect and vacuum polarization.

The equation (\ref{important1}) becomes essentially simpler if the
potential is close to the Coulomb one, and the deviation $\delta
V(r)=V(r)-V_C(r)$ can be treated as a perturbation. For the pure
Coulomb case, when ${\partial E_{\rm C}}/{\partial m} ={E_{\rm C}}/
{m}$, we immediately obtain the well-known result (see, e.g.,
\cite{OS,shabaev})
\begin{equation}\label{Coulomb0}
\Delta g_{\rm C} =
-\frac{\kappa}{2j(j+1)}\,\left[1-2\kappa\frac{\gamma +n_r}
{N}\right]\;.
\end{equation}
where
$$N=\sqrt{(\gamma+n_r)^2+(Z\alpha)^2}\, ,\quad
\gamma=\sqrt{\kappa^2-(Z\alpha)^2}\, ,$$
and $n_r$ is the radial quantum number.
The correction to the energy is
\begin{equation}
\delta E = \int{d^3r}\,\Psi^+(r) \delta V(r) \Psi(r)\;.
\end{equation}
As follows from the dimensional reasons, the wave function in the
Coulomb field can be presented in the form $\Psi(r)=m^{3/2}
\widetilde{\Psi}(mr)$, where $\widetilde{\Psi}$ is dimensionless.
Passing to the variable $\rho=mr$, we find
\begin{equation}
\delta E = \int{d^3\rho}\,\widetilde{\Psi}^+(\rho) \delta V(\rho/m)
\widetilde{\Psi}(\rho)\;.
\end{equation}
Taking the derivative over $m$ and returning to the variable $r$,
we obtain
\begin{equation}
\frac{\partial \delta E} {\partial m} = -
\frac{1}{m}\int{d^3r}\,\Psi^+(r) r \,\frac{\,\partial \delta
V(r)}{\partial r}\Psi(r)\;,
\end{equation}
and
\begin{equation}\label{important13}
\delta g =
-\frac{\kappa}{2j(j+1)}\,\left[1+\frac{2\kappa}{m}\bigg\langle r\,
\frac{\,\partial \delta V(r)}{\partial r} \bigg\rangle\right]\;.
\end{equation}

Eqs. (\ref{important1}), (\ref{important13}) are the basis of our
approach.

\section{Anomalous magnetic moment}

It may be interesting to generalize our results to the case of a
particle with a non-vanishing anomalous magnetic moment. In
particular, it is necessary for antiprotonic atoms. The anomalous
magnetic moment of the antiproton is big because of the complicated
internal structure of the particle. Therefore, it makes sense to
consider the modification of Eqs. (\ref{important1}),
(\ref{important13}) due to the anomalous magnetic moment separately
from the radiative corrections.

As known, the modified Dirac hamiltonian for a particle with the
anomalous magnetic moment in an arbitrary electromagnetic field is
of the form (see, e.g., \cite{BLP})
\begin{equation}\label{Dir_eq}
H=\gamma_0\,\bigl({\bm \gamma}\cdot({\bm p}+e{\bm A})\bigr) +
\gamma_0 m - eA_0+
\frac{i\mu_a}{2}\,\gamma_0\,\sigma_{\mu\nu}F^{\mu\nu}\;,
\end{equation}
where $\sigma_{\mu\nu}=(1/2)[\gamma_\mu,\gamma_\nu]$, $\mu_a$ is the
dimensional anomalous magnetic moment, $F^{\mu\nu}$ is the tensor of
the electromagnetic field; we remind that the charge of the bound
particle is $(-e)$.

As follows from Eq.~(\ref{Dir_eq}), the correction to the energy
of a bound particle in the homogenous magnetic field (\ref{defAB})
is of the form (cf.~Eq.~(\ref{eq:TM0}))
\begin{equation}
 \Delta E = \int d^3r \,\overline{\Psi}({\bm r})
 \left\{e\bigl({\bm
\gamma}\cdot{\bf A} \bigr) -\mu_a\bigl( {\bm \Sigma}\cdot{\bf B}
\bigr) \right\}\Psi({\bm r})\,,
\end{equation}
where $\Sigma_j=(i/2)\epsilon_{jkl}\sigma_{kl}$ is the spin
operator. The wave function $\Psi({\bm r})$ is an eigenfunction of
the hamiltonian (\ref{Dir_eq}) with $-eA_0(r)=V(r)$ and $A_i=0$.
Since the potential  $A_0(r)$ is spherically symmetric, the wave
function still has the form (\ref{Dirac}), and the result for the
$g$ factor reads (cf.~Eq.~(\ref{gf1f2}))
\begin{eqnarray}\label{gfactora}
g&=&\frac{2m}{j(j+1)}\times \biggl\{\kappa\int{dr}\,r^3
f_1\,f_2 \nonumber\\
&-&\frac{\mu_a}{2e} \left[ \int{dr}\,r^2
\bigl(f_1^2-f_2^2\bigr)-2\kappa \right] \biggr\} \;.
\end{eqnarray}

The radial wave functions $f_1$ and $f_2$ satisfy the system of
equations (cf.~\cite{BLP})
\begin{eqnarray} \label{Diracrad}
f_2^\prime + (E-V-m)f_1 +
\left(\frac{1-\kappa}{r}-\frac{\mu_a}{e}V^\prime\right)f_2 = 0\;,
\nonumber\\
f_1^\prime
-(E-V+m)f_2+\left(\frac{1+\kappa}{r}+\frac{\mu_a}{e}V^\prime\right)f_1
= 0\;.
\end{eqnarray}
Then we multiply the first equation by $r^3f_2$ and the second one
by $r^3f_1$, sum up the results and take the integral over $r$.
Integrating by parts the terms with the derivatives, we arrive at
the following relation (cf.~(\ref{identity}))
\begin{eqnarray}\label{identity1}
\int{dr}\,r^3 \,f_1\,
f_2&=&-\frac{1}{4m}\biggl[1-2\kappa\int{dr}\,r^2
\bigl(f_1^2-f_2^2\bigr) \nonumber\\
&-&\frac{2\mu_a}{e} \int{dr}\,r^3
V^\prime\,\bigl(f_1^2-f_2^2\bigr)\biggr]\, .
\end{eqnarray}
Using this formula and Eqs. (\ref{g0}) and (\ref{g0HE}), we present
the result for the $g$ factor in the form
\begin{eqnarray}\label{important12}
g&=&\frac{1}{2j(j+1)}\,
\biggl\{-\kappa(1+2a)\nonumber\\
&+& (2\kappa^2+a)\frac{\partial E} {\partial m}\nonumber\\
&-&a\frac{\kappa}{m} \int{dr}\,r^3
V^\prime\,\bigl(f_1^2-f_2^2\bigr) \biggr\}\; ,
\end{eqnarray}
where $a=-2m\mu_a/e$ is the dimensionless anomalous magnetic moment
(for a free particle $g=2(1+a)$). This result is obtained for an
arbitrary potential $A_0$ and is exact in the parameter $a$. We
emphasize that the radial wave functions $f_1$ and $f_2$ and the
binding energy $E$ depend on the anomalous magnetic moment because
of the equations (\ref{Diracrad}).

In the nonrelativistic approximation, Eq. (\ref{important12}) turns
into
\begin{eqnarray}\label{g_nr}
g&=&\frac{1}{2j(j+1)}\, \biggl\{ (1-2\kappa)(a-\kappa)
\nonumber\\
&-& (2\kappa^2+2\kappa a+a)\frac1{2m}\int{dr}\,r^3 V^\prime\,
f^2_{nr} \biggr\}\; ,
\end{eqnarray}
where $f_{nr}$ is the radial part of the nonrelativistic wave
function. This formula is valid even if $a\sim 1$.

It is interesting also to consider an expansion of the $g$ factor
(\ref{important12}) in the parameter  $a$ for arbitrary field
strength (when the nonrelativistic approximation is not valid). The
linear in this parameter term reads
\begin{eqnarray}\label{a1}
\delta g_a&=&\frac{a}{2j(j+1)} \biggl\{-2\kappa+\frac{\partial E} {\partial m}\nonumber\\
&-& \frac{2\kappa^2}{m}\frac{\partial}{\partial m}
\int{dr}\,r^2 V^\prime\, f_1 f_2
\nonumber\\
&-&\frac{\kappa}{m} \int{dr}\,r^3
V^\prime\,\bigl(f_1^2-f_2^2\bigr) \biggr\}\;.
\end{eqnarray}
Here we used the formula for the linear in  $\mu_a$ correction to
the energy,
 \begin{eqnarray}\label{dEa}
\delta E_a=\frac{2\mu_a}{e}\int{dr}\,r^2 V^\prime\, f_1 f_2\;,
\end{eqnarray}
and took into account that $\partial \mu_a/\partial m=0$. Let us
consider Eq.(\ref{a1}) for the pure Coulomb potential. Strictly
speaking, the equations (\ref{Diracrad}) have no sense for a pure
Coulomb field because of the terms $\propto 1/r^2$, which lead to
the phenomenon of falling to the center. It is a consequence of the
point-like  source of the field and is absent if finite nuclear size
is taken into account. However, for large quantum numbers which are
mostly interesting for the experiments, the correction due to the
finite nuclear size does not change essentially the result obtained
from Eq.(\ref{a1}) for a pure Coulomb field. Using the radial matrix
element from Ref.\cite{shabaev}, we obtain
\begin{eqnarray}\label{aC}
\delta g_a&=&\frac{a}{2j(j+1)} \biggl\{-2\kappa+\frac{\gamma+n_r} {N}
-\frac{\kappa(Z\alpha)^2}{N^2}\nonumber\\
&-&\frac{4\kappa^2(Z\alpha)^4[2\kappa(\gamma+n_r)-N]}{\gamma(4\gamma^2-1)N^4}
\biggr\}\;,
\end{eqnarray}
where $N$ and $\gamma$ are defined after Eq. (\ref{Coulomb0}). We
see that the correction to the $g$ factor due to the anomalous
magnetic moment has essentially more complicated dependence on
quantum numbers than the leading term (\ref{Coulomb0}).

\section{Finite-nuclear-size effect}

As an illustration of  efficiency of our method, let us consider  a
finite-nuclear-size correction to the bound-electron g factor. For
$1s$ state, this problem was solved analitically in
Ref.\cite{Savelii2000} in the nonrelativistic approximation
($Z\alpha\ll 1$), and numerically in Ref.\cite{Beier2000} for
arbitrary value of the parameter $Z\alpha$. In
Ref.\cite{Shabaev2002}, the result was obtained analytically in the
next-to-leading approximation in the parameter $Z\alpha$, and for
arbitrary states. The result of Ref.\cite{Shabaev2002} allows one to
describe well the correction to the $g$ factor up to $Z=20$. On the
other hand, in Ref.\cite{Shabaev1993} the approximate formulas for
the finite-nuclear-size  correction to the energy levels, $\delta
E_{fns}$, were obtained, which gives the result with a relative
error of about $0.2\%$ up to $Z=100$. Using Eq.(\ref{important1})
and known dependence of $E_{fns}$ on $m$, see \cite{Shabaev1993}, we
obtain
 \begin{equation}\label{FNS}
\delta g_{fns}=\frac{\kappa^2}{j(j+1)}\, \frac{\partial E_{fns}}
{\partial m}=\frac{\kappa^2(2\gamma+1)}{j(j+1)}\, \frac{E_{fns}}
{m}\;.
\end{equation}
Moreover, using our method we can calculate the radiative correction
to $\delta g_{fns}$  for $s_{1/2}$ state and  $p_{1/2}$ state (and
arbitrary $n_r$) coming from the effect of vacuum polarization.  For
these states, the finite-nuclear-size effect is the most
significant. The radiative correction $\delta E_{fns}$ to the energy
$E_{fns}$ was considered in detail in Ref.\cite{MST2004}. The
correction due to the vacuum polarization $\delta E^{VP}_{fns}$ can
be represented in the form
\begin{equation}\label{dEvp}
\delta E^{VP}_{fns}=E_{fns} \Delta^{VP}\,,
\end{equation}
where $E_{fns}$ was obtained in Ref. \cite{Shabaev1993}, and the
explicit analytical form of $\Delta^{VP}$ for $s_{1/2}$ state and
$p_{1/2}$ state is derived in Ref.\cite{MST2004}. Eq. (\ref{dEvp})
with $E_{fns}$ from Ref.\cite{Shabaev1993} and $\Delta^{VP}$ from
Ref.\cite{MST2004} provides the high accuracy of $\delta
E^{VP}_{fns}$ up to $Z=100$. However, the correction $\Delta^{VP}$
was calculated in \cite{MST2004} for the mass of the bound particle
$m$ being equal to the mass $M$ of the particle in the fermion loop.
Therefore, in order to use Eq. (\ref{important1}) we should know the
explicit dependence of $\Delta^{VP}$ on $m/M$ (we can set $M=m$ only
after the differentiation over $m$). The corresponding expression
for $\Delta^{VP}$ can be easily obtained following the derivation of
$\Delta^{VP}$ in \cite{MST2004}. For $Z\alpha\sim 1$, the main
contribution to $\Delta^{VP}$ is given by the logarithmically
enhanced term
\begin{equation}
\Delta^{VP}\approx\frac{2\alpha(Z\alpha)^2}{3\pi\gamma_1} \ln^2(m
R)\;,
\end{equation}
where $\gamma_1=\sqrt{1-(Z\alpha)^2}$, and $R$ is the nuclear
radius. This term is the same for $s_{1/2}$ and $p_{1/2}$ states and
comes from the distances $1/m,\,1/M\gg r\gg R$. Therefore, it
contains only the logarithmic dependence on the mass of the bound
particle, in contrast to the power-like dependence of $E_{fns}$.
Within the logarithmic accuracy, we have
\begin{equation}\label{ddE}
\frac{\partial}{\partial m} \delta E^{VP}_{fns}\approx \Delta^{VP}
\frac{\partial}{\partial m} E_{fns}=(2\gamma_1+1)\frac{\delta
E^{VP}_{fns}}m \;.
\end{equation}
Thus, for $Z\alpha\sim 1$, we obtain
  \begin{equation}\label{FNS1}
\delta g_{fns}^{VP}=\frac43(2\gamma_1+1)\frac{\delta
E^{VP}_{fns}}{m}\;.
\end{equation}
For $Z\alpha\ll 1$ the dependence of $\Delta^{VP}$ on $m/M$ is more
complicated and will be considered elsewhere.

\section{Conclusion}

We present a new effective approach to obtain various corrections to
the bound particle $g$ factor using the corresponding corrections to
the energy levels. It should be noted that this method is
appropriate for the perturbations of potential-like type. There are
also corrections which cannot be derived via this  approach. The
leading part of the non-potential contribution is due to the virtual
Delbr\"uck scattering and it was calculated in Ref. \cite{pra}.

As other applications of our method, we can suggest the Uehling
potential contribution to the $g$ factor of bound electrons and
muons. In muonic atoms it is one of the dominant effects, while in
electronic hydrogen-like atoms the accuracy is essentially higher
(see, e.g., \cite{eions}).

For muonic atoms, the Uehling-potential correction to the energy
levels  was calculated in the nonrelativistic approximation  for
certain levels in \cite{pusto}. The result exact in $Z\alpha$ was
obtained in \cite{cjp98} for ground state of hydrogen-like atom and
agrees with Eq. (\ref{important1}). The methods developed there can
be easily applied to arbitrary levels and thus, using
Eq.~(\ref{important1}), one can find the correction to the $g$
factor due to the Uehling potential.

Using the known corrections to the energy levels coming from the
high-order terms of vacuum polarization, we can obtain the
corresponding contribution to the g factor. One correction comes
from the Wichmann-Kroll potential (which accounts for the
higher-order in $Z\alpha$ terms in the induced charge density), see
Review \cite{BR1982}. Others are the second-order correction with
respect to the Uehling potential, and first-order correction with
respect to the radiative correction to the Uehling potential (of
order of $\alpha$).

\section*{Acknowledgements}
We are grateful to V.G. Ivanov for useful discussions. A.I.M. thanks
the Max-Planck-Institute for Nuclear physics, Heidelberg, and the
Max-Planck-Institute for Quantum Optics, Garching, for hospitality
during the visit. The work was supported in part by RFBR Grants No.
03-02-16510, 03-02-04029, and 03-02-16843, and DFG GZ: 436 Rus
113/769/0-1R.


\begin{thebibliography}{00.}

\bibitem{BLP} V. B. Berestetskii, E. M. Lifshitz, and L. P. Pitaevskii,
\emph{Relativistic quantum theory} (Pergamon Press, Oxford, 1982).

\bibitem{rose} M. E. Rose. {\em  Relativistic Electron Theory.\/}
John Wiley \& Sons, Inc., New York, London, 1961.

\bibitem{OS} S. A. Zapryagaev, Opt. Spect. {\bf 47}, 9 (1979).

\bibitem{shabaev} V. M. Shabaev, in: {\em Precision physics
of simple atomic systems\/} ed. by S. G. Karshenboim and V. B.
Smirnov (Springer, Berlin, Heidelberg, 2003), p. 97.


\bibitem{Savelii2000} S.~G.~Karshenboim, Phys. Lett. A {\bf 266}, 380 (2000).

\bibitem{Beier2000} T.Beier, Phys. Rep. {\bf 339}, 79 (2000).

\bibitem{Shabaev2002} D. A. Glazov and V. M. Shabaev, Phys. Lett. A {\bf
    297}, 408 (2002).

\bibitem{Shabaev1993}  V. M. Shabaev, J. Phys. B: At. Mol. Opt. Phys. {\bf
    26}, 1103 (1993).

\bibitem{MST2004} A.I.Milstein, O.P.Sushkov, I.S.Terekhov, Phys. Rev. A {\bf 69}, 022114 (2004).

\bibitem{pra} R. N. Lee, A. I. Milstein, I. S. Terekhov, and S. G. Karshenboim,
Phys. Rev. A {\bf 71}, 052501 (2005).

\bibitem{eions}
H.~H\"affner, T.~Beier, N.~Hermanspahn, H.-J.~Kluge, W.~Quint,
S.~Stahl, J.~Verd\'u, and G.~Werth, Phys. Rev. Lett. \textbf{85}, 5308 (2000);\\
J.~L.~Verd\'u, S.~Djekic, S.~Stahl, T.~Valenzuela, M.~Vogel,
G.~Werth, T.~Beier, H.-J.~Kluge, and W.~Quint, Phys. Rev. Lett.
\textbf{92}, 093002 (2004).

\bibitem{pusto} G.~E.~Pustovalov, Sov. Phys. JETP {\bf 5}, 1234 (1957); \\
  D.~D.~Ivanenko and G.~E.~Pustovalov, Adv. Phys. Sci. {\bf 61}, 1943 (1957)

\bibitem{cjp98} S.~G.~Karshenboim, Can. J. Phys. {\bf 76}, 169 (1998);
  JETP {\bf 89}, 850 (1999); misprints in some asymptotics were corrected
  in:
  S.~G.~Karshenboim, V.~G.~Ivanov and V.~M.~Shabaev, Can. J. Phys.
 {\bf 79}, 81 (2001); JETP {\bf 93}, 477 (2001).


\bibitem{BR1982}   E. Borie and G.A. Rinker, Rev. Mod. Phys. {\bf 54} , 67 (1982).

%
%
%
%
%
%
%
%
%
%


\end{thebibliography}
\end{document}